\documentclass[12pt]{article}
\usepackage{amsmath}
\usepackage{graphicx}

\usepackage{epsf}

\usepackage{subcaption}
\usepackage{pifont}
\usepackage{caption}
\usepackage{algorithm,algorithmic}
\usepackage{setspace}
\usepackage{adjustbox}
\usepackage{algorithm}
\usepackage{algorithmic}

\usepackage{color, ulem}
\usepackage{booktabs}
\usepackage[table,x11names]{xcolor}
\usepackage{multirow}
\usepackage[toc,page]{appendix}
\usepackage{fancyhdr}
\usepackage{lipsum}
\usepackage{geometry}
\usepackage[utf8]{inputenc}
\usepackage[english]{babel}

\usepackage[round,authoryear]{natbib}
\title{\textbf{A Differential Degree Test for Comparing Brain Networks}}
\author{Ixavier A. Higgins$^1$ \and Ying Guo$^1$ *   \and Suprateek Kundu$^1$ \and Ki Sueng Choi$^2$ \and Helen Mayberg$^2$}

\date{$^1$Emory University, Rollins School of Public Health\\  $^2$Emory University School of Medicine, Department of Psychiatry and Behavioral Science\\ $\ast$ yguo2@emory.edu}

\begin{document}
\baselineskip=22pt
\newdimen\jot \jot=5mm
\newcommand{\fnote}[1]{\mbox{$\mbox{}^{#1}$}}
\newcommand{\bigsp}{\mbox{$\; \; \; \; \;$}}
\newcommand{\half}{\mbox{$\frac{1}{2}$}}
\newcommand{\bm}[1]{ \mbox{\boldmath $ #1 $} }

\maketitle
\vskip 24pt
\singlespacing
{\noindent \bf Abstract: } Recently, graph theory has become a popular method for characterizing brain functional organization.  One important goal in graph theoretical analysis of brain networks is to identify network differences across disease types or conditions. Typical approaches include massive univariate testing of each edge or comparisons of local and/or global network metrics to identify deviations in topological organization. Some limitations of these methods include low statistical power due to the large number of comparisons and difficulty attributing overall differences in networks to local variations in brain function. We propose a novel differential degree test (DDT) to identify brain regions incident to a large number of differentially weighted edges across two populations. The proposed test could help detect key brain locations involved in diseases by demonstrating significantly altered neural connections. We achieve this by generating an appropriate set of null networks which are matched on the first and second moments of the observed difference network using the Hirschberger-Qi-Steuer (HQS) algorithm. This formulation permits separation of the network's true topology from the nuisance topology which is induced by the correlation measure and may drive inter-regional connectivity in ways unrelated to the brain function. Simulations indicate that the proposed approach routinely outperforms competing methods in detecting differentially connected regions of interest. Furthermore, we propose a data-adaptive threshold selection procedure which is able to detect differentially weighted edges and is shown to outperform competing methods that perform edge-wise comparisons controlling for the error rate. An application of our method to a major depressive disorder dataset leads to the identification of brain regions in the default mode network commonly implicated in this ruminative disorder.\\


\section{Introduction}
\doublespacing


In recent years, graph theoretical tools have become increasingly important in the analysis of brain imaging data.  In particular, evaluations of the associations between spatially distinct regions has led to valuable insights into the brain's organization in health and disease.  Functional connectivity (FC), which measures the coherence between neurophysiological time series \citep{Friston1994}, has been extremely valuable in identifying disease-induced modifications to cortical and subcortical communication.  In fact, altered cortical activity has been observed in major depressive disorder (MDD) \citep{Craddock2009, Drysdale2017}, Alzheimer's disease \citep{Stam2006}, and schizophrenia \citep{Liu2008, Rubinov2009}.  While Pearson correlation is a widely used FC measure, alternate association metrics such as partial correlations \citep{Wang2016}, mutual information \citep{Salvador2005}, and coherence \citep{Bassett2011} are finding favor.  Brain networks have become particularly important since the FC measures offer different perspectives on co-activation between brain regions, and many studies agree that psychiatric disorders and neurodegenerative diseases manifest as disruptions in local and global functional connectivity \citep{Pandya2012}.

Many methods exist for comparing brain networks and connectivity patterns across populations.  The earliest approach tests for group differences at each edge in the network \citep{Nichols2002}. For a network with N regions, this requires multiple testing corrections since N(N-1)/2 unique edges must be assessed.  Unfortunately, controlling the family-wise error rate or false discovery rate leads to a reduction in power to detect group differences at the edge level.  The sum of powered score (SPU) and adaptive sum of powered score (aSPU) tests \citep{Pan2014} leverage edge level differences to assess overall deviation in the networks \citep{Kim2015}.  While they can lead to high powered tests, there are practical difficulties in selecting the optimal tuning parameter.  Furthermore, the tests do not specifically identify edges, regions, or structures contributing to overall network differences, which leads to a loss in interpretability in the brain functional aspect.

Other approaches assume differences in brain connectivity result in large deviations in the network's topology. The Network Based Statistic (NBS)  \citep{Zalesky2010} and PARD \citep{ShuoChen} are useful for identifying collections of differentially weighted edges (DWEs) forming interconnected subcomponents, but have reduced exploratory value.  A key assumption in these methods is that altered edges form connected subnetworks \citep{Kim2015}.  However, if this assumption is violated, NBS is severely underpowered to detect differences in the networks \citep{Zalesky2010}. Furthermore, NBS can only detect collections of DWEs and nodes involved in the larger connected component, potentially missing small subcomponents of connected DWEs responsible for differences. Other methods \citep{Rudie2013, Wang2015} have focused on comparing graph metrics across networks, using two sample t-tests to test for differences. Unfortunately, these tests may often be underpowered to detect group differences \citep{Kim2014}, and there are doubts on the suitability of two-sample t-tests to compare some network metrics \citep{Fornito2010, Hayasaka2010}. Alternatively, nonparametric approaches utilize permutation tests \citep{Zalesky2010} or generate random networks \citep{Bullmore1999global} in order to construct distributions for network metrics of interest under the null hypothesis and then use these reference distributions to evaluate the significance of the observed network features. However, generating the appropriate null network is non-trivial. Existing approaches attempt to randomly rewire edges while preserving the degree distribution and the clustering coefficient \citep{Bansal2009, Maslov2002, Volz2004}.  
Unfortunately, the network generation schemes are sensitive to the desired network measure (see \cite{Fornito2013graph} for an overview) and may not provide a complete picture of the network differences reflected by alternate summary measures.

In this paper, we propose a Differential Degree Test (DDT) to identify brain regions that demonstrate significant between-group differences in neural connections. The proposed testing approach facilitates the comparison of brain networks across populations while bypassing the drawbacks of current methods. In particular, the test is based on the difference network where the edges represent the statistical significance of between-group differences. The observed difference network is compared against a set of null networks which are carefully constructed to maintain both the first and second moment characteristics of the observed difference network using the Hirschberger-Qi-Steuer (HQS) algorithm. Retention of the first and second moments is critical to preserving the nuisance topology of the observed difference network while annihilating intrinsic group structures of the observed network. Such a null network reflects the true topology induced by the correlation measure encoding the brain connectivity, without being overly sensitive to any particular network metric. In contrast, naive random networks may not be adequate, since these networks exhibit non-random topologies that are largely attributed to transitivity (associations between two regions is substantially influenced by other regions) induced by the correlation measure \citep{zalesky}. Thus, such random null networks do not provide a truly random assortment of network configurations nor are they guaranteed to replicate the nuisance structures in the observed network. Ideally, one would like to construct appropriate null networks that increase the detection of edge weight difference driven by differences in brain functionality while reducing the identification of artificial substructures.

We adapt the null network generation scheme of \cite{hirschberger} to replicate networks retaining only the nuisance structure in the difference network, which enables the separation of the network's true topology from the nuisance topology. We note that while \cite{zalesky} utilize HQS to examine the impact of nuisance topology on local network features, our approach is the first to use the HQS algorithm for the assessment of network differentiation across populations. We further propose an adaptive thresholding procedure to identify significant DWEs by comparing against the generated null difference networks using the HQS algorithm. Based on the thresholded difference adjacency matrix, the proposed Differential Degree Test (DDT) identifies nodes or brain regions that demonstrate a significantly higher number of DWEs as compared to the null distribution.

Through extensive simulations, we illustrate that the proposed method has greater power to detect differentially connected nodes across networks compared to standard multiple testing procedures, while also maintaining reasonable control over false positives. Furthermore,  the adaptive threshold selection procedure leads to increased power to detect DWEs across the network as compared to Bonferroni and false discovery rate (FDR) correction procedures. Additionally, the adaptive threshold approach under the proposed method can automatically adapt to different network settings and hence is more generalizable compared to `hard' thresholding approaches assuming a fixed threshold. Finally, we apply the proposed approach for the analysis of a major depressive disorder (MDD) dataset, which leads to meaningful findings regarding disrupted brain connectivity due to MDD.

The rest of the paper is organized as follows. In sections \ref{sect:  method}, we discuss the construction of null difference networks and the proposed DDT procedure.  We present simulation results in Section \ref{sect: simulation}, MDD data analyses in Section \ref{sect: real data}, and conclude with a  discussion of the findings in Section \ref{sect:  discussion}.

\section{Method}\label{sect:  method}

In the following, we discuss the construction of subject-specific functional brain networks, formulation of the difference network, and the details of the proposed DDT. Our analysis investigates connectivity disruptions across the entire brain, but is also amenable to hypothesis-driven investigations of functional connectivity containing a number of pre-selected brain regions.

 \subsection{Brain network construction }\label{sect: network construction}
 In network analysis of neuroimaging data, the brain can be represented as a graph defined by a finite set of nodes (brain regions) and edges showing the statistical association between pairs of nodes. For $N$ nodes, the network is represented as a symmetric $N \times N$ connectivity matrix, $\boldsymbol{\mathrm{G}}$, which can be thresholded to obtain the adjacency matrix $\boldsymbol{\mathrm{A}}$, representing the edge set of the network. For selection of the node system, the naive approach is to treat each voxel as a putative region of interest. This approach results in an extremely high-dimensional connectivity matrix that not only poses challenges for subsequent analyses, but also tends to be unreliable and noisy. A more common approach is to define nodes based on anatomically defined brain structures, e.g. Automated Anatomical Labeling (AAL) atlas \citep{Tzourio2002} and Harvard-Oxford atlases \citep{Fischl2004, Frazier2005}.  When analyzing brain functional networks, it is suggested to parcellate the brain into putative functional areas based on clusters of voxels exhibiting similar signals in resting-state functional imaging data \citep{Craddock2012}.  Some widely used examples of functionally defined node systems are the Power 264 node system \citep{Power2011},  Yeo \citep{Yeo2011}, and Gordon \citep{Gordon2014} atlases, among others.

For brain network based on functional magnetic resonance imaging (fMRI), the edges represents the coherence in the temporal dynamics between the blood oxygen-level dependent (BOLD) signal between node pairs. In this paper, we utilizes undirected measures of connectivity such as Pearson and partial correlation, where Pearson correlation measures the marginal association between two regions and partial correlation measures their association conditioned on all other regions in the network.  Given the heavy debate on the merits and disadvantages of each correlation measure in brain network analysis \citep{Kim2015, Liang2012}, we investigate both and compare the findings.

 The resulting network, $\boldsymbol{\mathrm{G}}$, is a weighted graph representing undirected statistical associations between all pairs of nodes. Often, a thresholding procedure is applied to produce a binary adjacency matrix, {\bf A}, where a value of 1 in the $(i,j)^{th}$ entry indicates a connection between the respective regions.  This network formulation is particularly advantageous as it simplifies calculations of graph metrics and leads to intuitive metric definitions (see \cite{bullmore, rubinov} for more details). 

Since we are interested in between-group differences in functional networks, we  consider a difference network which is defined on the same node system as the functional network but the edges represent the strength of between-group differences in the functional connections. Details of the difference network construction are presented in the following section. We focus on the number of thresholded edges incident to each region in the difference network, which we call the differential degree. Similar to the interpretation of nodal degree in connectivity matrices, we focus upon this metric as it suggests regions contributing to local differences in the network architecture across diseases or conditions. We believe that a brain region incident to a large number of differentially weighted edges (DWEs) is potentially responsible for overall differences in brain network topology, without being sensitive to any particular network summary measure commonly used to capture connectome differences.




\subsection{Differential Degree Test}
In this section, we present the DDT method for identifying brain nodes whose connections demonstrate significant differences between group.

\subsubsection{Difference Network Construction}
Suppose we are comparing networks between two groups with  $n_{r}$ subjects in group $r$ ($r=1,2$). Denote $\boldsymbol{G}^{k_r}=\{g_{r,ij}^{k_{r}}\}$ ($k_{r}=1,\dots ,n_{r}$) as the estimated brain connectivity matrices for the $k_r$th subject in the $r$th group ($r=1,2$) and $ g_{r,ij}^{k_{r}}$ denotes the connectivity measure (such as the Pearson or partial correlation) between nodes $i$ and $j$ ($i,j \in \mathcal{N}=\{1,\ldots, N\}$) for the $k_r$-th subject in the $r$th group.  The first step of DDT is to construct a $N \times N$  difference network $\boldsymbol{\mathrm{D}}=\{d_{ij}: i,j \in \mathcal{N}\}$, where $d_{ij}$ represents the statistical significance of population-level differences in the connection strength between node $i$ and $j$, i.e.
  \begin{equation} \label{diff_network}
     d_{ij}= 1-p(\{g_{1,ij}\}, \{g_{2,ij}\}) \in [0,1),
			\end{equation}

\noindent where $p(\{g_{1,ij}\}, \{g_{2,ij}\} )$ is the p-value of a between-group difference test based on the estimated connectivity measures at edge $(i,j)$ across subjects in the two groups. For example, one can obtain the p-value by applying two-sample t test to  $\{g_{r,ij}^{k_1}\}$ and $\{g_{r,ij}^{k_2}\}$ . We will provide more detailed discussion on how to derive the p-values from various types of between-group tests in section \ref{sect:pval}.  From (\ref{diff_network}), each element in the difference network $ d_{ij}$ serves as our measure of the difference of the edge connectivity $g_{r,ij}$ between the two groups, with larger values (i.e. smaller p-values) corresponding to larger group differences at the $(i,j)$th edge, and vice-versa.  Note that $\boldsymbol{\mathrm{D}}=\{d_{ij}\}$ is a symmetric matrix where $\forall i,j \in \mathcal{N}$ $d_{ij}=d_{ji} $ and $d_{ij}=0$  for $i=j$ given that we are not interested in the diagonal elements.

From the difference network $\boldsymbol{\mathrm{D}}=\{d_{ij}\}$, we can derive the difference adjacency matrix {\bf A}$=\{a_{ij}\}$ where $a_{ij}$ represent the presence of group differences in the connection between nodes $i$ and $j$, i.e.

  \begin{equation} \label{eq:tau}
     a_{ij}= I(d_{ij} >  \tau),
  \end{equation}

\noindent where $ \tau$ is a threshold for selecting edges which are differentially weighted. When $d_{ij}$ exceeds the threshold $\tau$, or equivalently the p-value for the group test is smaller than $1- \tau$, we obtain $a_{ij}=1$ indicating the presence of group difference at the edge $(i,j)$. Otherwise, $a_{ij}=0$ represents no group difference at the edge $(i,j)$.  In the following section, we will present a data-driven adaptive threshold selection method for finding $\tau$.

Based on the difference adjacency matrix $\boldsymbol{\mathrm{A}}$, we define the following differential degree measure for the $i$th node ($i=1,\ldots, N$),
   \begin{equation} \label{degree_eq}
   d_{i}=\sum_{j \in \mathcal{N}, j \ne i} a_{ij}
   \end{equation}

The differential degree measure  $d_{i}$ represents the number of connections to node $i$ that demonstrate significant differences between the two groups  as captured by edge-wise p-values without multiplicity adjustments. In subsequent steps of the DDT, $d_{i}$ will be used as the test statistic for investigating node $i$'s contribution to disrupted communication in the brain. While the difference network provides edge-level information on between group differences, it is widely accepted that cognitive deficits in mental diseases  are demarcated by disruptions in systems \citep{Catani2005}. Thus, collections of connected DWEs are more consistent with the system wide disruption paradigm than evaluation of individual DWEs. The DWEs incident to each node form a locally connected component and indicate that irregular activity at the node of interest contributes to differentiated co-activation with adjacent regions.  Investigation at the nodal level not only has biological justification, but also substantially improves the multiple testing problem. The number of statistical tests scales linearly with the network's size rather than quadratically at the edge level. The notion of disruptions in sub-systems has also been used in previous work to mitigate the multiplicity problem common to network comparisons \citep{Zalesky2010}.

\subsubsection{Deriving p-value from between-group tests}\label{sect:pval}
The p-value used to define the difference network in (\ref{diff_network}) can be derived based on various between-group testing procedures. The p-values fall into two categories: model-free and model-based. The model-free p-values are derived based on parametric or nonparametric tests between the two groups of subjects without accounting for the subjects' biological or clinical characteristics. The common choices of such tests include the two-sample t test, the nonparametric Wilcoxon rank sum test or the permutation test.  The model-based p-values are derived from regression models where the subject-specific connectivity measure (or some transformation) is modeled in terms of group membership and other relevant factors such as age and gender that may affect the brain connectivity.  These p-values for between-group difference can then be derived based on the test of the parameter in the model associated with the group covariate. This model-based p-value reflects the degree of group differences while controlling for potential confounding effects.  In many neuroimaging studies, subjects' group memberships are not based on randomization but rather based on observed characteristics. In this case, the distribution of subjects' demographic and clinical variables tend to be unbalanced between the groups and there often exist some potential confounding factors in between group comparisons  \citep{Satterthwaite2015}. For such studies, it may not be the case that the model-based p-values more accurately reflect group-induced variation in functional connectivity as compared with model-free p-values.

We note that when computing the difference network in (\ref{diff_network}), the proposed approach does not apply a multiple testing correction to the edge-wise between-group test p-values. Such multiplicity adjustment often reduces the power to detect DWEs. Additionally, since our goal is to detect differentially expressed nodes in the brain network, a multiplicity adjustment on the edge-wise tests is not crucial, provided the falsely identified DWEs are more or less uniformly distributed across the nodes without systematic differences. In such a case, the threshold $\tau$ in (\ref{eq:tau}), which is chosen using an appropriately constructed null distribution as in Section 2.2.4, automatically adjusts for falsely identified DWEs occurring across nodes. Indeed, extensive simulation studies where the proposed method is able to control false positives at a nominal value.

\subsubsection{Null distribution generation}
After constructing the difference network $\boldsymbol{\mathrm{D}}=\{d_{ij}\}$ and deriving the differential degree measure, $d_i$, for each node, the next step in the DDT procedure is to conduct a statistical test to evaluate whether there is significant group difference in the connections to the node. As a standard strategy in hypothesis testing, we will evaluate the test statistic,  $d_i$, with respect to its null distribution under the hypothesis that there are no between-group differences. For this purpose, we first derive the null distribution by generating difference networks under the null hypothesis.

We present a procedure for generating null difference networks that maintain some of the fundamental characteristics of the observed difference networks but has a random pattern of between-group differences which is expected under the null hypothesis.  Since the elements in the difference network lie within a restricted range, i.e. $(0,1)$, we first apply a logit transformation, i.e.
\begin{equation}
\bar{\boldsymbol{\mathrm{D}}}= \{\bar{d}_{ij}: \bar{d}_{ij}=logit(d_{ij})\in (-\infty, \infty), i<j, i,j \in N\}.
\end{equation}

We define the first and second moment characteristics for the observed difference network as follows,
\begin{equation*}
\bar{e}=E[\bar{d_{ij}}]\ \text{and} \ \bar{v}=Var[\bar{d_{ij}}] \ \text{for i$<$j} \text{, }\quad e=E[\bar{d}_{ij}] \quad \text{for i$=$j}
\end{equation*}
 where $\bar{e}$ represents the mean of the off-diagonal elements, $e$ represents the mean of the diagonal element and $\bar{v}$ is the variance of the off-diagonal elements.

In the following, we present a procedure for generating a null difference network $\boldsymbol{\mathrm{C}}\in \rm I\!R ^{NxN}$ whose first and second moment characteristics matches that of the observed difference network, and preserves its true topology.  Motivated by the Hirshberger Qi-Steuer (HQS) algorithm \citep{hirschberger}, we propose to generate $\boldsymbol{\mathrm{C}}$ based on the multiplication of a random matrix and its conjugate transpose
\begin{equation}
  \boldsymbol{\mathrm{C=L*L}}^{T},
\end{equation}

\noindent where $\boldsymbol{\mathrm{L}}= \in R^{N \times m}$.  Based on the formulation of \citep{hirschberger}, we generate $l_{ij} \sim N(\mu,\sigma^2)$   where $\mu=\sqrt{\frac{\bar{e}}{m}}$ and $ \sigma^2=-\mu^2 + \sqrt{\mu^4+\frac{\bar{v}}{m}}$ and $m=min\{2, \lfloor \frac{e^2 - \bar{e}^2}{v}\rfloor\}$ where $\lfloor \cdot \rfloor$ is the floor function.  Based on this specification, we can show that
\begin{center}
 $E[c_{ij}]=\bar{e}$, $Var[c_{ij}]=\bar{v} $ and $E[c_{ii}]=e$,
\end{center}
 Please see equations \ref{appendix:expectation} and \ref{appendix:variance} for details. The generated null difference network $\bar{\boldsymbol{\mathrm{C}}}$ maintains the first and second moment characteristics of the observed difference network $\bar {\boldsymbol{\mathrm{D}}}$.  Finally, we transform $\bar{\boldsymbol{\mathrm{C}}}$ through the inverse logit function to obtain a null difference network {\bf C} such that $c_{ij} \in (0,1)$.

The proposed generation procedure has several appealing features. First, it is a very fast algorithm for generating null networks. Second, the generated null difference network, $\bar{\boldsymbol{\mathrm{C}}}$, preserve the first and second moment characteristics of the observed difference network $\bar{\boldsymbol{\mathrm{D}}}$.  An important advantage in maintaining these fundamental properties of the observed network is that it will help make the generated null network a meaningful reference for comparison with the observed network. For example, to perform meaningful comparison of the connectivity structure between two networks, a critical condition is that the two networks must have similar number of edges \citep{deVico}. This condition would be violated if there exists a significant difference in the average connectivity measure between the two networks in the sense that the network with higher average connectivity is associated with larger number of edges. By generating null networks with the same first and second moment as the observed network, the proposed procedure makes sure the comparison between the observed network against the null networks would not be confounded by the their differences in the fundamental characteristics. More importantly, replication of the first and second moments allows the null networks to preserve the nuisance topology of the observed difference network while annihilating intrinsic group structures of the observed network. As discussed in \cite{zalesky}, benchmarking against such null networks permits identification of the intrinsic topology in the observed network.

\subsubsection{An adaptive threshold selection method}\label{sect:thresholding}

Recall that after obtaining the difference network $\bar{\boldsymbol{\mathrm{D}}}=\{\bar{d}_{ij}\}$, we need to threshold it to derive the difference adjacency matrix ${\boldsymbol{\mathrm{A}}}=\{\bar{a}_{ij}\}$. If $\bar d_{ij} > \gamma$, $\bar a_{ij}=1$ indicating the presence of a group difference at the edge $(i,j)$ where $\gamma=logit(\tau)$. Otherwise, $\bar a_{ij}=0$ represents no group difference at the edge $(i,j)$.

In the existing between-group network tests, the threshold value is typically selected by a multiple comparison method that controls the family-wise error rate or the false discovery rate.  Others select a pre-specified cutoff or grid over a range of cutoffs \citep{Zalesky2010}. We propose to adaptively select the threshold based on the distribution of the between-group test statistic.  Specifically, the $\tilde{c}_{ij}$ are independent and identical samples from the mixture distribution,H(.),

\begin{equation}
\label{mixdist}
\mathrm{H}(\tilde{c}_{ij}) = \frac{2\sigma^2}{4}T-  \frac{2\sigma^2}{4}Q, \\
\end{equation}
where $T$ and $Q$ are non-central $\chi^2 $ and central $\chi^2$ random variables, respectively.  Each variable in the mixture distribution depends only on the mean and variance of the observed data (see Appendix \ref{appendix:HQS}).  We propose two ways to select the threshold, $\gamma$ as the 95$^{th}$ quantile: (1) aDDT which uses the theoretical critical value based on the parametric mixture distribution in (\ref{mixdist}), and (2) eDDT which uses the empirical critical  value based on the empirical distribution. The numerical advantages and disadvantages of each of the two thresholding methods will be addressed in the simulation studies. Since the null difference network, $\bar{\boldsymbol{\mathrm{C}}}$, is generated in a way that it matches the first and second moments of the observed difference network, $\bar{\boldsymbol{\mathrm{D}}}$, the selected threshold value $\gamma$ will automatically adapt to the properties of the observed difference network. Compared to hard thresholding approaches which use a fixed cut-off value, our threshold selection method can potentially provide an adaptive and general approach for choosing suitable threshold values for different studies. Once the threshold value $\gamma$ is computed as above, one can apply it to the generated null difference networks $\bar{\boldsymbol{\mathrm{C}}}$ to obtain difference adjacency matrices $\bar{\boldsymbol{\mathrm{A}}}=\{\bar{a}_{ij}\}$ such that $\bar{a}_{ij}=1$ if $\bar{c}_{ij}>\gamma$ and $0$ otherwise.

The proposed threshold selection procedure controls the selection of false positive edges, while circumventing the loss of power inherent in existing multiplicity corrections methods. This is achieved by adaptively selecting the threshold based on the distribution of the elements in the difference network. Similar approaches \citep{Newton2004, Kundu2018} have effectively controlled type I error by using the empirical distribution of edge probabilities to select a threshold in order to detect important connections.

\subsubsection{The DDT Test}

In this section, we present a statistical test for the difference degree measure, $d_i$, for node $i$ based on the generated null difference networks.  Recall that $d_i=\sum_{j \in \mathcal{N}, j\ne i} a_{ij}$ where $a_{ij}$ is a binary variable indicating the presence of group difference at the connection between node $i$ and $j$. $\bar{d}_i$ essentially is a count variable representing the number of connections out of a total of $N-1$ connections of node $i$ that show between group difference. Therefore, we can model ${d}_i$ with a binomial distribution. Under the null, ${d}_i \sim f^{null}=\textrm{Binomial}(N-1, p_{i}^{null})$  where $p_{i}^{null}$ is the the expected probability for each connection of node $i$ to demonstrate between group difference under the null hypothesis. We can estimate the null probability $p_{i}^{null}$ based on the generated null difference networks, that is,

\begin{equation*}
\label{p_null}
\hat{p_{i}}^{null}=\frac{1}{M(N-1)}\sum_{m=1}^{M}\sum_{j \in \mathcal{N}, j\ne i} \bar{a}_{ij}^{(m)},
\end{equation*}
where M is the total number of null networks and $\bar{a}_{ij}^{(m)}$ are elements of $m^{th}$ thresholded null network, $\bar{\boldsymbol{\mathrm{A}}}^{(m)}$. By comparing the observed ${d}_i$ against the null distribution, we identify all regions incident to more DWEs than is expected by chance.

In the following algorithm, we summarize the procedure of the DDT.

\begin{algorithm}
\caption{DDT Procedure}\label{alg:euclid}
\begin{algorithmic}[1]
\STATE Construct the difference network $\bar{\boldsymbol{\mathrm{D}}}=\{\bar{d}_{ij}\}$ where $\bar{d}_{ij}=logit(1-p(\{g^{k1}_{ij}\}, g^{k2}_{ij}\}))$
\STATE Obtain the first and second moment of $\bar{\boldsymbol{\mathrm{D}}}$, i.e. $\bar{e}=E(\bar{d}_{ij})$ and $\sigma^{2}=Var(\bar{d}_{ij})$
\STATE Generate M null Difference Networks  $\bar{\boldsymbol{\mathrm{C}}}^{m}(m=1,\ldots,\mathrm{M})$ based on $\bar{e}$ and $\sigma^2$ using the proposed procedure such that $\bar{\boldsymbol{\mathrm{C}}}^{m}$ preserves the first and second moment characteristics of $\bar{\boldsymbol{\mathrm{D}}}$
\STATE Apply the adaptive threshold selection method to find $\gamma$ based on the percentile of the non-central Chi-square distribution of the null difference networks. Specifically, aDDT uses parametric percentiles and eDDT uses empirical percentiles in the adaptive thresholding.
\STATE Apply threshold  $\gamma$ to $\bar{\boldsymbol{\mathrm{D}}}$ to obtain the difference adjacency matrix ${\boldsymbol{\mathrm{A}}}=\{\bar{a}_{ij}\}$ and the difference degree measure ${d}_i$ for node $i \in \mathcal{N}$
\STATE Derive the null distribution for ${d}_i$, i.e. $f^{null}$ based on the generated null difference networks.  Specifically, $ f^{null} =\mathrm{Binomial(N-1, p_{i}^{null})}$, where $p_{i}^{null}$ is obtained from $\{\bar{A}^{m}\}$ based on equation \ref{p_null}.
\STATE Compare the observed difference degree measure, ${d}_i$, against its null distribution, $f^{null}$, to evaluate the significance of between-group differences in the connections of node $i$.
\end{algorithmic}
\end{algorithm}

\section{Simulation}\label{sect:  simulation}

We conduct extensive simulation studies to assess the proposed method's ability to detect regions with significantly different connections between two groups of subjects. Unless otherwise noted, the generated networks contain $N=35$ nodes,  and we consider sample sizes of $20$ and $40$ for each of the two groups. For the first set of simulations, we consider the case where there is only one node in the network incident to a specified number of DWEs. Without loss of generality, we refer to it as node 1, and assess whether the proposed DDT can accurately identify this node. We consider both DDT methods, i.e. aDDT based on parametric percentiles and eDDT based on empirical percentiles in the adaptive thresholding step. Three network structures are considered in the simulation: (1) random (2) small world (3) hybrid. Random networks contain edges that are equally likely to be positive or negative for all connections.  We generate this structure by sampling edge weights independently from a $N(0,.04)$ distribution, which produces a connectivity matrix with no structural zeros. The small world network retains the cliquishness of the regular lattice and the short path length of the random network. This structure retains small world properties observed in human brain networks \citep{Hilgetag2016}.  The hybrid network seeks to fuse the block diagonal structure observed in real brain networks, while maintaining the small world-ness inherent to human brains.  The ``blocks" correspond to functional modules observed in the brain such as the default mode and visual networks.

In our simulations, all subjects share a common base brain network, $\boldsymbol{\mathrm{B}}$, which is a correlation matrix generated according to the random, small world and hybrid network structure. We perturb the edge weights in {\bf B} to induce subject-level network variation and control the distribution of DWEs across the populations. For subjects $i_1=1,...,n_{1}$ and $i_2=1,...,n_{2}$ in the two groups, we generate the subject-level networks, {\bf H}$_{i_1}$ and {\bf H}$_{i_2}$, as follows: for $n_1$ subjects in group 1, {\bf H}$_{i_1}=\boldsymbol{\mathrm{B + W}}_{i_1}$, where $\boldsymbol{\mathrm{W}}_{i_1}\in R^{N\times N}$, $w_{ij, i_1} \sim N(0,.02) \quad \text{for } 1 \le i<j\le N=35$ and $w_{ij,i_1}=0$ $\forall i=j$; for $n_2$ subjects in population two, {\bf H}$_{i_2}=\boldsymbol{\mathrm{B + W}}_{i_2}$ where {\bf W}$_{i_2}\in R^{N\times N}$. Let I be the set of differentially connected nodes where I$=\{1\}$ for the first set of simulation. For $i\in$I, we generate $q$ off-diagonal elements corresponding to the DWEs in the $i$-th row and column edges connected with $i$ from $N(.1,.02)$ and other edges of $i$ from $N(0,.02)$. For $i\not\in$I, we have $w_{ij,i_2}\sim N(0,.02)$. We consider $q=$4, 7 and 11 to assess our method's power to detect differentially connected region(s) when the number of DWEs increases.  We construct the difference network with model-free p-values, where we conduct a two sample t-test and record one minus the p-value as the weight for each edge. 

We compare the performance of DDT to that of two other tests.  The first comparison method ($\mathrm{T_{(10\%)}}$) is a standard two sample t-test of local nodal degree.  For this test, we threshold the subject-specific correlation matrices to attain 10\% density, evaluate the subject-level degree measure at each node and then perform a two sample t-test to compare the nodal degree across groups. We also investigated but did not include the results obtained from 15\% density and 1\% network density, which were less powerful in detecting differentially connected regions than 10\% density. We also consider two binomial tests which are similar to DDT in that they directly assess the number of differentially weighted edges incident to a node but differ from DDT in that they apply some multiple comparison corrections to detect the DWEs. Specifically, the first binomial test, $\mathrm{Bin_B}$, applies a Bonferroni correction to detect the DWEs \citep{Tyszka} and the second binomial test, $\mathrm{Bin_F}$, implements a less stringent FDR multiple testing correction. For both binomial tests, each node's differential degree is the sum of all DWEs incident to it. We do not compare to NBS since it assesses global structures among DWEs whereas we are interested in local topological differences.

In the second set of simulations, we assess the methods' performances when there are 3 differentially connected regions. We consider two scenarios in this setting. First, the network size is fixed while the number of DWEs varies with $q=$ 4, 7 and 11. Second, we fix the proportion of DWEs for the differentially connected nodes to be 30\% while increasing the size of the network. We report various metrics to quantify the methods' accuracy in detecting differentially connected nodes across the simulations.  The false positive rate(FPR) is calculated as $\sum_{s=1}^S\sum_{n=1}^{N}I(\hat{R}_{n,s}=1 , R_n=0) / (S*N)$ and quantifies the chance that each method incorrectly identifies a differentially connected region. The true positive rate (TPR) is calculated as $\sum_{s=1}^S\sum_{n=1}^{N}I(\hat{R}_{n,s}=1 , R_n=1) / (S*N)$) and measures the correct identification. Here, $S$ is the total number of simulations. $\hat{R}_{n,s}$ takes the value 1 if region $n$ in simulation $s$ is selected as differentially connected and 0 otherwise.  $R_n$ is a binary indicator of whether region $n$ is differentially connected in the ground truth. We compare accuracy in selecting truly differentially connected regions by Matthews correlation coefficient (MCC) \citep{Johnstone2012}, which is a popular measure for accessing the correspondence between predicted and true class labels. MCC, which is computed as $\small \frac{TP\times TN-FP\times FN}{\sqrt{(TP+FP)(TP+FN)(TN+FP)(TN+FN)}}$, takes values in $[-1, 1]$ where 1 indicates perfect agreement between the predicted and true class labels, 0 no agreement, and -1 inverse agreement. In this formula, $TP, TN, FP,FN$ denote the number of nodes that are true positives, true negatives, false positives and false negatives, respectively. In a supplementary analysis of the simulation results, we assess the performance of the adaptive thresholding procedures presented in section \ref{sect:thresholding} in correctly detecting DWEs. We compare the MCC in selecting the true DWEs based on the proposed aDDT and eDDT thresholding procedures with that based on two hard thresholds at .95 and .99 as well as based on multiple comparison corrections thresholds using the Bonferroni and FDR methods.

\subsection{ Results}
Table \ref{table:Simulation 1 results} displays accuracy measures for identifying one differentially connected node across two populations in the first set of simulations. Generally, the proposed DDT methods, i.e. aDDT and eDDT,  exhibit larger TPR than the T-tests and the Binomial tests across various sample sizes and network structures. The Binomial tests achieve the lowest FPR, which is attributed to the Bonferroni and FDR multiple testing corrections.  However, the multiplicity corrections reduce the power to detect the correct region. The T-test attains the nominal type I error rate ($\alpha=.05$). For all methods, the TPR improves when the sample size increases and the number of differentially connected edges increase. Overall, the two proposed DDT approaches exhibit superior performance as compared to the other tests. Among the two DDT methods,  $\mathrm{eDDT}$ typically exhibits higher TPR, but the latter has a slightly higher FPR, although the FPR under both approaches is less than the nominal level of .05.

\begin{figure}[ht!]
 \centering
 \caption{{\small Comparison of eDDT, aDDT, t-test (T$_{(10\%)}$), binomial tests (Bin$_F$, Bin$_B$) in the second set of simulations with three differentially connected nodes incident to four ({\it first row}), seven ({\it second row}), and eleven ({\it third row}) DWEs. The true positive rate (TPR), false positive rate (FPR) and Matthew's correlation coefficient (MCC) is presented for all methods across the three network structures considered and the red dashed line demarcates the nominal significance level (.05). DDT exhibits superior performance in detecting the differentially connected nodes while not exceeding the allowable type I error rate. }}
 \mbox{\includegraphics[height=5in,width=5.5in]{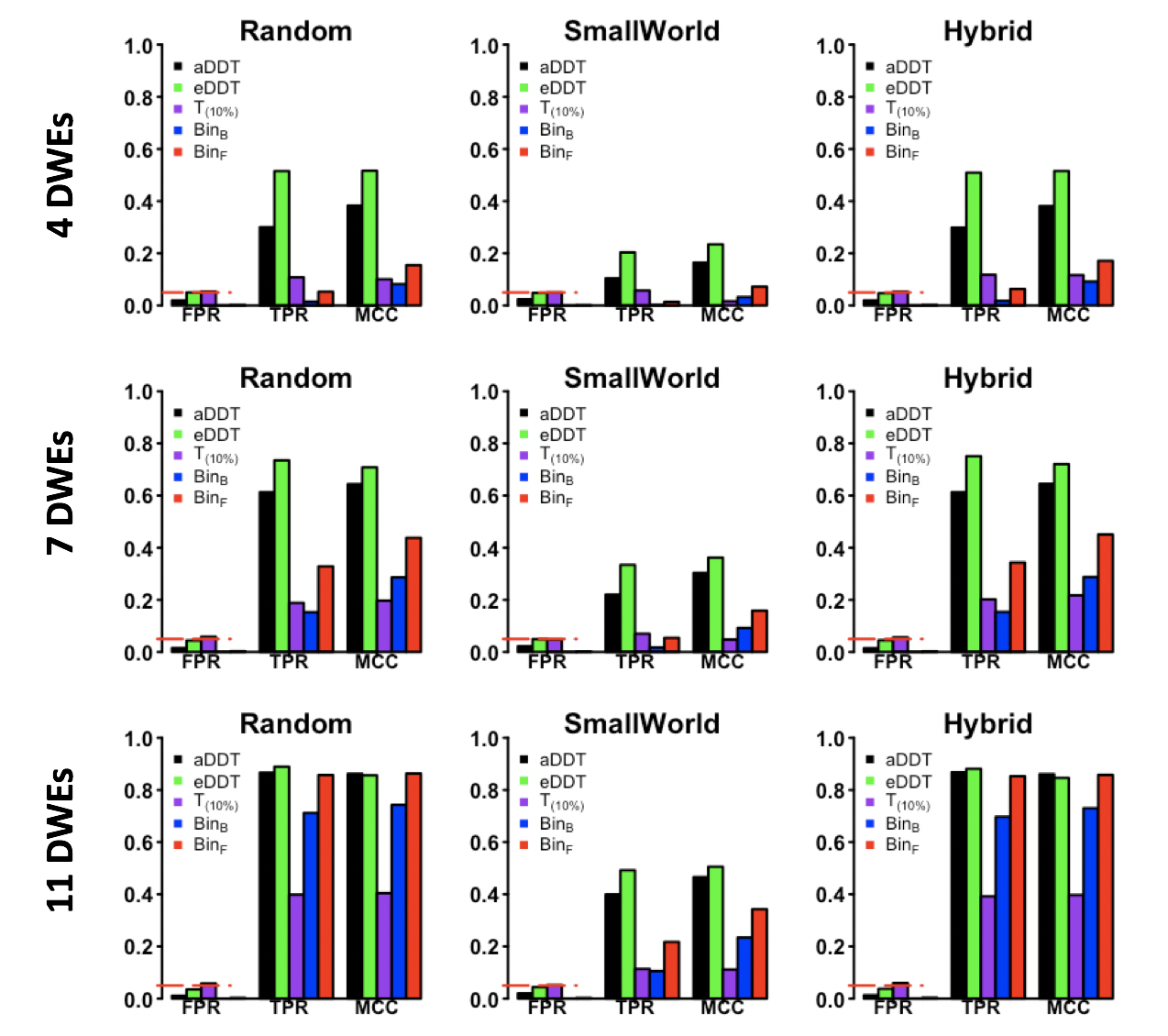}} \\
\label{fig:threeDCN}
\end{figure}

\begin{table}[h!]
\caption{False positive and true positive rates for the Random, Small World, and Hybrid data designs considered.   }
\label{table:Simulation 1 results}
\resizebox{1\textwidth}{!}{
\begin{tabular}{cccccccccccccc}
\toprule\hline  \\[-.7em]
&&  &\multicolumn{5}{c}{\large{\textbf{n$_{1}=$n$_2=$20}}}  & &  \multicolumn{5}{c}{\large{\textbf{n$_1=$n$_2=$40 }}}   \\  \midrule

   &  &  & \multicolumn{2}{c}{{\textbf{DDT}}}  &\multicolumn{2}{c}{{\textbf{Binomial}}} & {\textbf{T-test}}  &&\multicolumn{2}{c}{{\textbf{DDT}}}  &\multicolumn{2}{c}{{\textbf{Binomial}}} & {\textbf{T-test}}\\
&  \textbf{Network Structure}  &  \textbf{DWE*}    &    \multicolumn{1}{c}{\bf{$\mathrm{aDDT}$}}         & \multicolumn{1}{c}{\textbf{$\mathrm{eDDT}$}}         &     \multicolumn{1}{c}{\textbf{$\mathrm{Bin_B}$}}             &    \multicolumn{1}{c}{\textbf{$\mathrm{Bin_F}$}}  & \textbf{10\% }  &  &\multicolumn{1}{c}{\textbf{$\mathrm{aDDT}$}}         & \multicolumn{1}{c}{\textbf{$\mathrm{eDDT}$}}         &     \multicolumn{1}{c}{\textbf{$\mathrm{Bin_B}$}}             &    \multicolumn{1}{c}{\textbf{$\mathrm{Bin_F}$}}  & \textbf{10\%}    \\
       \cline{4-8} \cline{10-14}\\[-.9em]
\multirow{3}{*}{\rotatebox[origin=l]{90}{{\textbf{FALSE POSITIVE RATE}}}}
 & {\textbf{Random}}                  &       4       &  .021 & .046  &.002 & .001 & .052   &&.019 &.046  &.002   &.002 & .052 \\
   &                 &          7       & .022  & .047  &.002 & .002 & .051   && .017 &.044  &.002   &    .002  & .054 \\
   &                 &         11       & .020  & .046  &.002 & .002 & .052   &&.014 &.044  &.003   & .002  & .055 \\
   &                 &         20       & .017  & .045  &.0004 & .003 & .057   &&.008 &.033  &.0008   & .004  & .060 \\[.2cm]
  & {\textbf{Small world}} &       4       & .022 & .046  &.001 & .001 & .050   && .019 &.045  &.002   & .002  & .054 \\
   &                 &          7       &  .023 & .046  &.002 & .002 & .051   && .016 &.044  &.002   & .002  & .051 \\
   &                 &         11       &  .020 & .045  &.002 & .002 & .052   && .012 &.042  &.003   &  .002 & .055 \\
     &                 &         20       & .019  & .045  &.0004 & .003 & .054   &&.008 &.033  &.0005   & .003  & .058 \\[.2cm]
  & {\textbf{Hybrid}}                 &       4       & .024  & .042  &.001 & .002 & .054   && .019 &.045  &.001  & .001   & .055 \\
   &                                          &          7    & .022 & .045  &.002 & .001& .055   && .018 &.045  &.002   & .002  & .059 \\
   &                                          &         11    & .020 & .046  &.002 & .002 & .054   && .015  &.046 &.002  & .002   & .064 \\
     &                                        &         20       & .017  & .045  &.0005 & .003 & .058   &&.009 &.034  &.0006   & .003  & .063 \\[.1cm] \\[-1.1em]
       \rule{0pt}{5ex}
   \multirow{3}{*}{\rotatebox[origin=l]{90}{{\textbf{TRUE POSITIVE RATE}}}}
 & {\textbf{Random}}                  &       4       &  .370 & .458  &.036 & .049 & .111   &&.710 &.891  &.123   &  .112    & .203 \\
   &                                             &          7    & .631  & .767  &.240 & .292 & .230   &&.977 &.991  &.747   &  .738    & .379 \\
   &                                             &         11    &  .893 & .885  &.694 & .686& .450   &&1.00 &.999  &.999   &  .998    & .619 \\
   &                                             &         20   &  .994 & .991  &.981 & .999 & .791   &&1.00 &1.00  &1.00   &  1.00    & .962 \\[.2cm]
  & {\textbf{Small world}}                  &       4       &  .287 & .505  &.040 & .034& .131   && .699 &.913  &.113   & .123 & .227 \\
   &                                                   &       7       & .639  & .738  &.274 & .226 & .155   && .982  &.994  &.784   & .764 & .307 \\
   &                                                   &       11     &  .895 & .908  &.696 & .692 & .551   && 1.00 &.999  &.998   & .999 & .840 \\[.2cm]
   &                                             &         20   &  .994 & .991  &.981 & .999 & .791   &&1.00 &1.00  &1.00   &  1.00    & .962 \\[.2cm]
  & {\textbf{Hybrid}}                 &       4       & .284  & .423  &.042 & .052 & .255   && .650 &.884  &.136   & .114      & .435 \\
   &                                          &       7       & .571  & .675  &.225 & .222 & .524   && .974 &.995  &.707   & .719      & .650 \\
   &                                          &       11     & .874  & .888  &.693 & .693 & .681   && .999 &1.00  &.997   & . 994     & .739 \\
   &                                          &       20     & .996  & .988  &.975 & .993 & .820   && 1.00 &1.00  &1.00   & 1. 00     & .978 \\[.3cm] \hline\bottomrule

   \end{tabular}
}
 \scriptsize*Number of Differentially Weighted Edges incident to node 1\\
  \scriptsize $\boldsymbol{\mathrm{Bin_B}}=$Binomial, bonferroni correction; $\boldsymbol{\mathrm{Bin_F}}=$ Binomial, FDR correction \\
   \scriptsize $\boldsymbol{\mathrm{aDDT}}=$DDT, theoretical threshold; $\boldsymbol{\mathrm{eDDT}}=$ DDT, empirical threshold \\
\end{table}
The advantages of proposed $\mathrm{aDDT}$ and $\mathrm{eDDT}$ over the alternative methods persist in the second set of simulations where three regions are differentially connected. In Figure \ref{fig:threeDCN}, across all network structures with a fixed number of nodes (N=35) and with four, seven, or eleven DWEs incident to each of the three nodes of interest, the DDT methods have the highest power to detect the regions of interest while attaining FPR comparable to that of $\mathrm{T_{(10\%)}}$. We note that our method is superior to the multiplicity corrected Binomial tests when the differentially connected regions are incident to a small to moderate number of DWEs and is comparably powered to detect differentially connected nodes as the FDR corrected tests when the number of DWEs is large. Furthermore, as in the first simulation setting, $\mathrm{eDDT}$ typically exhibits higher TPR than $\mathrm{aDDT}$, but the former has slightly higher FPR compared to aDDT. Notably, both methods exhibit FPR values close to the nominal level of 0.05.

\begin{figure}[ht!]
 \centering
 \caption{{\small  Performance of aDDT, eDDT, t-test($\mathrm{T_{10\%}}$), and binomial ($\mathrm{Bin_B}$, $\mathrm{Bin_F}$) tests in identifying the differentially connected node as the network size increases and the proportion of DWEs is fixed at 30\%. (Note, the results of the $\mathrm{Bin_B}$, $\mathrm{Bin_F}$ tests are very close and hence the two lines overlap)}}

\mbox{\includegraphics[height=2.5in,width=5in]{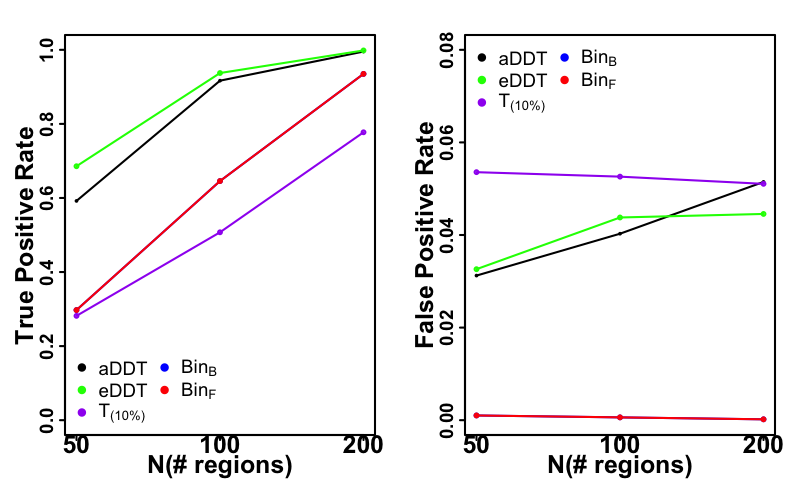}} \\
\label{fig:Rand3R_increaseN}
\end{figure}

We also examine at the performance of the approaches as the number of nodes increases, while keeping the proportion of DWEs incident to the region of interest fixed at 30\%.  Figure \ref{fig:Rand3R_increaseN} clearly illustrates the advantages of DDT for detecting regions incident to DWEs, while having a comparable or lower FPR as the network's size increases. Consistent with Table \ref{table:Simulation 1 results} and Figure \ref{fig:threeDCN}, $\mathrm{eDDT}$ exhibits the best TPR while the multiplicity corrected binomial tests have the smallest FPR, although the FPR levels under the DDT approaches are less than or equal to the nominal level across varying numbers of nodes. However, the TPR for eDDT and aDDT becomes increasingly similar as the number of regions is increased.

\begin{figure}[ht!]
 \caption{\small{Comparison of thresholding procedures implemented in aDDT and eDDT as well as four competitors (.95 and .99 hard threshold; FDR and Bonferroni multiplicity corrections) in detecting differentially weighted edges.}}
 \label{ThreshEffectiveness}
        \centering
        \includegraphics[width=.5\linewidth]{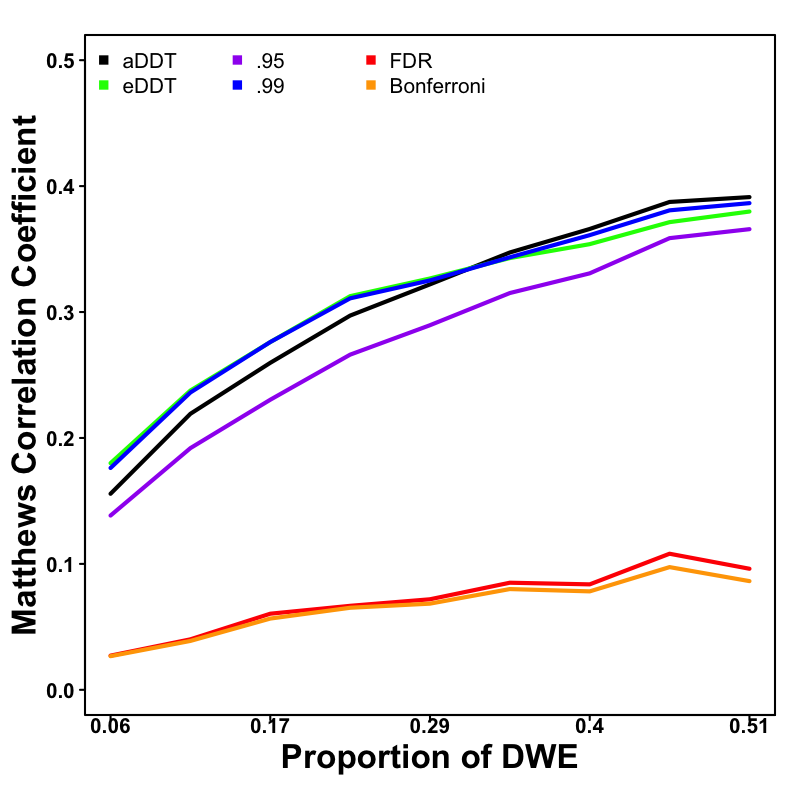}
\end{figure}

Although detection of regions incident to a significant number of DWEs is our primary focus, we also investigate the performance of the thresholding procedure fir detecting DWEs in a supplementary analysis in terms of the MCC values. Figure \ref{ThreshEffectiveness} indicates that aDDT's and eDDT's adaptive thresholding procedures outperform the Bonferroni and FDR multiplicity corrections over varying proportion of DWEs. Moreover, our method also exhibits superior MCC than the arbitrary hard threshold of 0.95, and at least one of the aDDT and eDDT approaches perform as well as the conservative hard threshold set at 0.99 as the proportion of DWEs across the network increases.

 \section{Data Application}\label{sect:  real data}
Existing literature has identified multiple brain regions implicated in major depressive disorder (MDD).  For example, MDD patients experience reduced connectivity in the fronto-parietal network as well as modified activity in areas such as the  insula \citep{Deen2010}, amygdala \citep{Sheline1998}, hippocampus \citep{Lorenzetti2009, Schweitzer2001}, dorsomedial thalamus \citep{Fu2004, Kumari2003}, subgenual and dorsal anterior cingulate cortex \citep{Mayberg1999}.  We apply the DDT to a MDD resting-state fMRI study \citep{Dunlop2017} to investigate brain regions contributing to differences in overall functional network organization in the affected population.

To construct brain network, we choose the 264-node system defined by \citep{Power2011}. Each node is a 10mm diameter sphere in standard MNI space representing a putative functional area consistently observed in task-based and resting-state fMRI meta-analysis. We focus upon 259 nodes located in cortical and subcortical regions, excluding a few nodes lying in the cerebellum. Each node is assigned to one of twelve functional modules defined in \cite{Power2011}: sensor/somatomotor (SM), cingulo-opercular task control (CIO), auditory (AUD), default mode (DMN), memory retrieval (MEM), visual (VIS), fronto-parietal task control (FPN), salience (SAL), subcortical (SUB), ventral attention (VAN), dorsal attention (DAN), and uncertain (UNC).

We measure the functional association between all pairs of brain regions with Pearson correlation and partial correlation.  Partial correlations are estimated using the {\it DensParcorr} R package \citep{Wang2016}. 
For both correlation measures, we conduct the between-group tests on the Fisher Z-transformed correlation coefficient at each edge and derive both the unadjusted p-values when confounding variables are not accounted for and also the adjusted p-values when they are accounted for.

We construct the difference networks based on the model-free and model-based between-group test p-values for connectivity measured by both Pearson correlation and partial correlation. We apply the proposed eDDT method to identify brain regions incident to a statistically significant number of DWEs. Subsequently, we investigate the distribution of the DWEs across the networks as well as between and within functional modules. 

 \subsection{Data preprocessing}
The data consists of resting-state fMRI scans from twenty MDD subjects and nineteen healthy subjects.  MDD patients are on average 45.8 years old (SD: 9.6 years) and fifty percent male.  The matched healthy participants are 47\% male and 43 years old (SD: 8.9 years). MDD patients were evaluated with the 17-item, clinician-rated Hamilton Rating Scale for Depression and had Mean(SD)of 19(3.4) which corresponds to severe depression \citep{Brown2008}.

During rs-fMRI scans, participants were instructed to rest with eyes closed without an explicit task. Data was acquired on a 3T Tim Trio MRI scanner with a twelve-channel head array coil.  fMRI images were captured with a z-saga sequence to minimize artifacts in the medial PFC and OFC due to sinus cavities \citep{Heberlein2004}.  Z-saga images were acquired interleaved at 3.4x3.4x4 mm resolution in 30 4-mm thick axial slices with the parameters FOV=220x220 mm, TR=2920 ms, TE=30 ms for a total of 150 acquisitions and total duration 7.3 min. Several standard preprocessing steps were applied to the rs-fMRI data, including despiking, slice timing correction, motion correction, registration to MNI 2mm standard space, normalization to percent signal change, removal of linear trend, regressing out CSF, WM, and 6 movement parameters, bandpass filtering (0.009 to 0.08), and spatial smoothing with a 6mm FWHM Gaussian kernel.

\subsection{Results}

Table  \ref{tab1:Top 20 Differentially Connected Nodes}(A) and \ref{tab1:Top 20 Differentially Connected Nodes}(B) list the top twenty differentially connected nodes for model-based Pearson and partial correlations.  Pearson correlation generally leads to more DWEs incident to nodes.  Thirty percent of the regions identified in Table \ref{tab1:Top 20 Differentially Connected Nodes}(A) are located in the SM module while twenty percent are in the DMN. Similarly, Table \ref{tab1:Top 20 Differentially Connected Nodes}(B) shows DMN nodes are extremely prominent (35\%) as well as the FPN and CIO which compose the task control system.   Jointly, these results suggest that altered connectivity in the DMN differentiates the brain networks in the MDD population from healthy controls.

\begin{table}[ht!]
\centering
\caption{Top twenty differentially connected nodes in the major depressive disorder study based on (a) model-based Pearson correlations and (b) model-based partial correlations. } \label{tab1:Top 20 Differentially Connected Nodes}
\begin{adjustbox}{width=.8\textwidth, height=.55\columnwidth}
\begin{tabular}{llllll}\hline \hline
\multicolumn{6}{l}{\bf (A) model-based Pearson correlations} \\ \hline
{\textbf{X}} & {\textbf{Y}} &{\textbf{Z}} & {\textbf{Name}} &{\textbf{Module}} & {\textbf{\#DWE}} \\ \hline
-53 & -22 & 23 & SupraMarginal\_L (aal)      & Auditory  & 29  \\
52 & 7 & -30 & Temporal\_Pole\_Mid\_R (aal) & Default mode & 27 \\
46 & 16 & -30 & Temporal\_Pole\_Mid\_R (aal) & Default mode & 26 \\
29 & 1 & 4 & Putamen\_R (aal) & Subcortical & 26 \\
47 & -30 & 49 & Postcental\_R (aal) & Sensory/somatormotor Hand & 23 \\
10 & -46 & 73 & Precuneus\_R (aal) & Sensory/somatormotor Hand & 19 \\
-24 & -91 & 19 & Occipital\_Mid\_L & Visual & 19 \\
 -54 & -23  & 43 & Parietal\_Inf\_L  & Sensory/somatomotor Hand & 18 \\ 
 31 &33 & 26 & Frontal\_Mid\_R & Salience & 18\\ 
 51 & -29 & -4 & Temporal\_Mid\_R (aal) & Ventral attention & 18 \\ 
 13 & -33 & 75 & Postcentral\_R (aal) & Sensory/somatomotor Hand & 16 \\ 
 -46 & 31 & -13 & Frontal\_Inf\_Orb\_L (aal) & Default mode & 16\\ 
 23 & 10 & 1 & Putamen\_R (aal) & Subcortical & 15\\ 
 -44 & 12 & -34 & Temporal\_Pole\_Mid\_L (aal) & Default mode & 15\\
 31 & -14 & 2 & Putamen\_R (aal) & Subcortical & 15 \\
 -38 & -27 & 69 & Postcentral\_L (aal) & Sensory/somatomotor Hand & 14\\
 -60 & -25 & 14 & Temporal \_Sup\_L (aal) & Auditory & 14 \\
 27 & 16 & -17 & Insula\_R (aal) & Uncertain & 14 \\
 52 & -2 & -16 & Temporal\_Mid\_R (aal) & Default mode & 14 \\
 50 & -20 & 42 & Postcentral \-R (aal) & Sensory/somatomotor Hand & 13 \\ \hline \hline
\multicolumn{6}{l}{\bf(B) model-based partial correlations} \\ \hline
-31                            & 19                             & -19                            & Frontal\_Inf\_Orb\_L (aal)        & Uncertain                           & 15                                 \\
24                             & 32                             & -18                            & Frontal\_Sup\_Orb\_R (aal)        & Uncertain                           & 13                                 \\
-38                            & -15                            & 69                             & undefined                         & Sensory/somatomotor Hand            & 13                                 \\
-26                            & -40                            & -8                             & ParaHippocampal\_L (aal)          & Default mode                        & 13                                 \\
-31                            & -10                            & -36                            & Fusiform\_L (aal)                 & Uncertain                           & 13                                 \\
17                             & -91                            & -14                            & Lingual\_R (aal)                  & Uncertain                           & 12                                 \\
-16                            & -46                            & 73                             & Parietal\_Sup\_L (aal)            & Sensory/somatomotor Hand            & 11                                 \\
23                             & 33                             & 48                             & Frontal\_Sup\_R (aal)             & Default mode                        & 11                                 \\
-28                            & -79                            & 19                             & Occipital\_Mid\_L (aal)           & Visual                              & 11                                 \\
37                             & -81                            & 1                              & Occipital\_Mid\_R (aal)           & Visual                              & 11                                 \\
-42                            & -55                            & 45                             & Parietal\_Inf\_L (aal)            & Fronto-parietal Task Control        & 11                                 \\
-54                            & -23                            & 43                             & Parietal\_Inf\_L (aal)            & Sensory/somatomotor Hand            & 10                                 \\
7                              & 8                              & 51                             & Supp\_Motor\_Area\_R (aal)        & Cingulo-opercular Task Control      & 10                                 \\
-45                            & 0                              & 9                              & Rolandic\_Oper\_L (aal)           & Cingulo-opercular Task Control      & 10                                 \\
-60                            & -25                            & 14                             & Temporal\_Sup\_L (aal)            & Auditory                            & 10                                 \\
-13                            & -40                            & 1                              & Precuneus\_L (aal)                & Default mode                        & 10                                 \\
-68                            & -23                            & -16                            & Temporal\_Mid\_L (aal)            & Default mode                        & 10                                 \\
-10                            & 39                             & 52                             & Frontal\_Sup\_Medial\_L (aal)     & Default mode                        & 10                                 \\
22                             & 39                             & 39                             & Frontal\_Sup\_R (aal)             & Default mode                        & 10                                 \\
-8                             & 48                             & 23                             & Frontal\_Sup\_Medial\_L (aal)     & Default mode                        & 10    \\ \hline
\end{tabular}
\end{adjustbox}
\end{table}

\begin{figure}[h!]
 \caption{\small{Differentially weighted edges detected by eDDT in the major depressive disorder study under the four difference network configurations: (a) model-free Pearson and (b) model-based Pearson (c) model-free partial (d) model-based partial.  Red edges indicate the average edge weight in the MDD population is statistically smaller than in healthy adults whereas blue edges demarcate the edge is statistically larger.  The network is decomposed into 12 functional modules: Sensor/somatomotor (SM), Cingulo-opercular Task Control (CO), Auditory( Aud), Default Mode (DMN), Memory Retrieval (Mem), Visual (Vis), Fronto-parietal Task Control (FP), Salience (SN), Subcortical (Sub), Ventral attention (VAN), Dorsal attention (DAN), Uncertain (Un).}}
 \label{fig:DifferenceNetworks_overall}
    \centering
     \includegraphics[height=5in,width=5in]{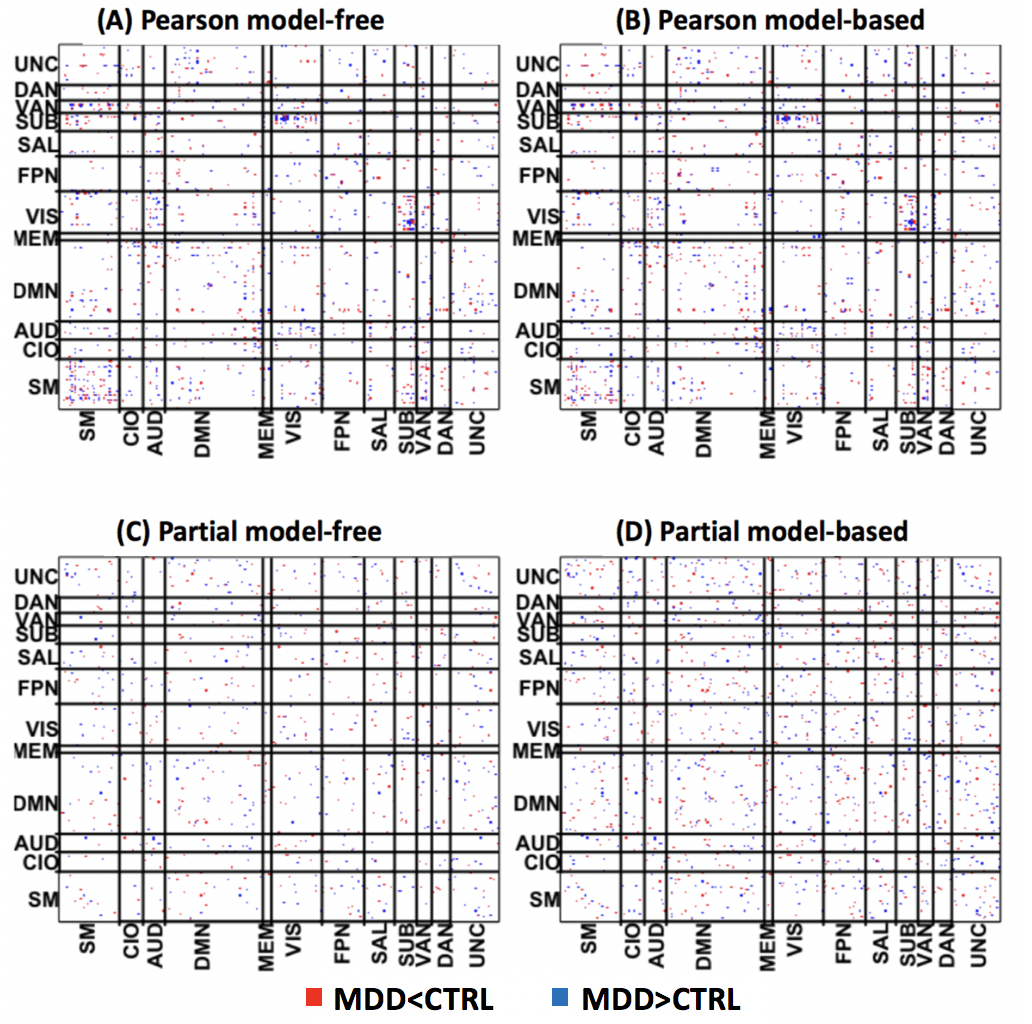}
\end{figure}

Figure \ref{fig:DifferenceNetworks_overall} displays the distribution of DWEs across the respective difference network. Here, we group the nodes based on the functional module assignment provided in \cite{Power2011}. The diagonal blocks represent within-module connections while the off-diagonal blocks represent between-module connections. For Pearson model-free and model-based analyses, we identified 793 and 776 DWEs, respectively. For partial correlations, we identified 458 DWEs based on model-free p-values and 772 DWEs for model-based p-values. The Pearson correlation derived difference networks exhibit spatial clustering of DWEs, specifically within the SM and between the SUB and VIS functional modules.  Table \ref{tableEdgecompare} reports the consistently and inconsistently detected DWEs when comparing the four difference networks investigated, i.e. model-free/model-based Pearson correlation networks and model-free/model-based partial correlation networks. Insignificant edges persist across all the difference networks considered and account for as much as  90\% of the edges in the networks. Generally, the findings are more consistent between the model-free and model-based p-values within the same correlation measure and less consistent across correlation measure.

\begin{table}[h!]
\caption{\small{Consistency of DWEs identified based on four difference networks. Values presented in the table are the number of network edges for each results consistency classification.}}
\label{tableEdgecompare}
\begin{center}
\begin{adjustbox}{width=1\textwidth, height=.65\columnwidth,keepaspectratio}
\begin{tabular}{llcccc}
\hline
 \multicolumn{2}{c}{Difference Network} & Sig in I& Insig in I & Sig in I & Insig in I \\
\multicolumn{1}{c}{I} & \multicolumn{1}{c}{II} & Sig in II & Insig in II & Insig in II & Sig in II \\ \hline \hline
Pearson model-free & Pearson model-based & 662 & 32504 & 131 & 114 \\
Pearson model-free & Partial model-free & 51 & 32392 & 691 & 406 \\
Pearson model-based & Partial model-based & 73 & 31936 & 703 & 699 \\
Partial model-based & Partial model-free & 454 & 32936 & 4 & 318 \\
\hline
{\footnotesize The total number of edges in the network is 33411.}
\end{tabular}
\end{adjustbox}
\end{center}
\end{table}

The distribution of DWEs within and between functional modules provides insight into disrupted communication among functionally segregated sub-systems in the brain. We conduct analysis to identify functional modules that are associated with higher number of DWEs as compared with other modules. Specifically, we propose the following chi-square statistic to help identify functional module pairs for which there are unusually high number of DWEs than what is expected by chance,

\begin{equation}
\label{chi}
X^{2}_{g_1,g_2}=\frac{(Q_{(g_1,g_2)}-E_{(g_1,g_2)})^{2}}{E_{(g_1,g_2)}},
\end{equation}
where $g_1 \in \{1,\ldots,\mathcal{G}\}$ and $g_2 \in \{1,\ldots,\mathcal{G}\}$ are indices corresponding to one of the $\mathcal{G}=12$ functional modules.  When $g_1=g_2$, $(g_1, g_2)$ represents a within module block, whereas it represents a  between-module block when $g_1\ne g_2$.  $Q_{(g_1,g_2)}$ represents the observed number of DWEs in the $(g_1, g_2)$ block and $E_{(g_1,g_2)}$ represents the expected number of DWEs in the $(g_1, g_2)$ block when the edges distribute randomly across the module blocks in the network. Let $|g|$ represent the total number of nodes within the $g$th module, and $p^{\ast}$ represent the proportion of DWEs among all the edges across the network.  It is straightforward to see that $E_{g_1,g_2}=p^{\ast}[\frac{|g_1|*(|g_2|-1)}{2}]$ for within module blocks, i.e. $g_1=g_2$, and $E_{g_1,g_2}=p^{\ast}*[|g_1|*|g_2|]$ for between-module blocks.

\begin{figure}[!h]
 \centering
 \caption{{\small Heat map of the $X^2_{(g1,g2)}$ statistic for (A) model-free Pearson correlations and (B) model-based Pearson correlations.  Red squares indicate modules with more statistically significant DWEs than would be expected by random chance.  We control the overall false discovery rate by only selecting module pairs with a multiplicity corrected p-value $<$ .05 . }}
\mbox{\includegraphics[height=3in,width=4in]{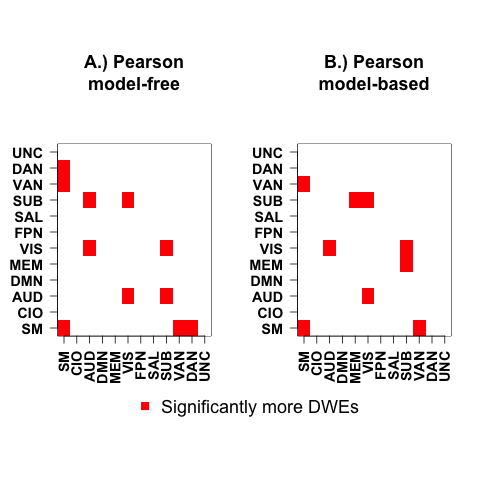}} \\
\label{fig:invertedchisq}
\end{figure}

Figure \ref{fig:invertedchisq} displays functional modules and module pairs exhibiting a significantly high number of DWEs based on the thresholded chi-square test statistic. The results are derived from the model-free Pearson correlations (Figure \ref{fig:invertedchisq}(A)) and model-based Pearson correlations (Figure \ref{fig:invertedchisq}(B)), respectively. Based on model-free Pearson correlations (Figure \ref{fig:invertedchisq}(A)), there are significantly high number of DWEs within the sensorimotor module and between the module pairs of sensorimotor-ventral attention, sensorimotor-dorsal attention, visual-auditory, subcortical-auditory and subcortical-visual. After accounting for age and gender, the model-based Pearson correlations (Figure \ref{fig:invertedchisq}(B)) also exhibit a large number of DWEs within the sensorimotor module and between the module pairs of sensorimotor-ventral attention, visual-auditory and subcortical-visual. However, the model-based Pearson correlations no longer show significantly high number of DWEs between the sensorimotor-dorsal attention and subcortical-auditory module pairs. Instead, the model-based correlations find significant number of DWEs between the subcortical-memory module pair which is not identified by the model-free Pearson correlations.

 \pagebreak

\clearpage
\section{Discussion}\label{sect:  discussion}

While the estimation of brain networks is gaining increasing attention in the neuroimaging literature, the fundamental question of how brains differ in functional organization across disease populations is not yet resolved. Our proposed method exhibits two strengths. First, our automated threshold selection permits identification of DWEs without sacrificing power as is the case with many methods dependent upon multiplicity corrections. Second, we use the generated null networks to test if each brain region is incident to more DWEs than would be expected by random chance.

We hypothesize that network wide dysconnectivity is driven by brain regions that irregularly communicate with other regions. The results from the real data analysis suggest that the DDT appropriately identifies problematic brain regions in major depressive disorder.  The existence of differential connectivity between nodes in the auditory and visual networks (Figure \ref{fig:invertedchisq}) has previously been observed \citep{Eyre}. Further, multivariate pattern analyses have suggested that the most discrimative functional connectivity patterns lie within and across the visual network, DMN, and affective network \citep{Zeng2012}.  The parahippocampal gyrus, which we detect as a problematic region (Table \ref{tab1:Top 20 Differentially Connected Nodes}), has also suggested as a region with differentiated connectivity patterns in depressed populations.   

Our simulation results demonstrate superior performance of the proposed DDT tests.  Although the binomial test's false positive rate is smaller than DDT, its ability to detect differentially connected nodes is severely attenuated.  DDT maintains the desired false positive rate while achieving higher true positive rates than the t-test across all network structures and sample sizes considered. Simulations indicate DDT's adaptive threshold selection is superior to conservative FDR and Bonferroni adjustments.

An obvious limitation to this work is the \textit{i.i.d.} assumption on edge weights in the null networks.  Although the independence assumption did not severely impact the method's performance relative to suitable competitors, it is likely that incorporation of inter-edge dependence structures will lead to better power to detect differentially connected nodes.

\begin{appendices}
\section{}\label{appendix:HQS}
\subsection{Proof for HQS procedure}
\setcounter{equation}{0}
\renewcommand{\theequation}{\thesection.\arabic{equation}}
In section 2.2.3, we suggest that sampling $l_{ij} \sim$ N($\mu$,$ \sigma^2$) appropriately allows for the condition that that $E(c_{ij})=\bar e$ and $Var(c_{ij})=\bar v$.  We now provide details for the distribution of $c_{ij}=\sum_k^m l_{ik}\times l_{kj}$.  Consider $l_{ik}$,$l_{kj} $$\overset {i.i.d.} \sim$ N($\mu, \sigma^2$) for $k=1,..,m$.
\begin{equation}
\label{eqn:summation}
\begin{split}
\sum_{k=1}^m l_{ik} l_{kj} & = \frac{1}{4}\sum_{k=1}^m(l_{ik} + l_{kj})^2 -  \frac{1}{4}\sum_{k=1}^m(l_{ik} - l_{kj})^2\\[-5mm]
\end{split}
\end{equation}
Note that for $l_{ik}$, $l_{k,j}$ $\overset {i.i.d.} \sim$ N($\mu, \sigma^2$),
\begin{equation*}
\begin{split}
& \frac{(l_{ik}+l_{kj})}{\sqrt{2}\sigma} \sim N(\frac{2\mu}{\sqrt{2}\sigma},1) \implies \frac{(l_{ik}+l_{kj})^2}{2\sigma^2} \sim \chi_1^2(\frac{4\mu^2}{2\sigma^2})\\[-5mm]
& \frac{(l_{ij}-l_{kj})}{\sqrt{2}\sigma}\sim N(0,1)\implies  \frac{(l_{ij}-l_{kj})^2}{2\sigma^2} \sim \chi_1^2\\[-5mm]
\end{split}
\end{equation*}

We can introduce constants and rewrite (\ref{eqn:summation}) as
\begin{flalign}\nonumber
\sum_{k=1}^m l_{ik} l_{kj} & = \frac{1}{4}\sum_{k=1}^m(l_{ik} + l_{kj})^2 -  \frac{1}{4}\sum_{k=1}^m(l_{ik} - l_{kj})^2 \nonumber\\[-5mm]
\sum_{k=1}^m l_{ik} l_{kj} & =\frac{2\sigma^2}{4}\sum_{k=1}^m\frac{(l_{ik} + l_{kj})^2}{2\sigma^2} -  \frac{2\sigma^2}{4}\sum_{k=1}^m\frac{(l_{ik} - l_{kj})^2}{2\sigma^2} \nonumber\\[-5mm]
\sum_{k=1}^m l_{ik} l_{kj} & = \frac{2\sigma^2}{4}T-  \frac{2\sigma^2}{4}Q
\end{flalign}

\noindent where T is a non-central $\chi^2$ with m df and non-centrality parameter $m\times(\frac{4\mu^2}{2\sigma^2}$) and Q is a central $\chi^2$ with m df.  Utilizing the first moment of non-central $\chi^2$ and central $\chi^2$ distributions, we see that
\begin{equation}\label{appendix:expectation}
\begin{split}
E[\frac{2\sigma^2}{4}T-  \frac{2\sigma^2}{4}Q] &= \frac{2\sigma^2}{4}E[T] -  \frac{2\sigma^2}{4}E[Q]\\[-5mm]
&= \frac{2\sigma^2}{4}(m+\frac{4m\mu^2}{2\sigma^2}) -  \frac{2\sigma^2}{4}m\\[-5mm]
&= m \mu^2 = \bar e\\[-5mm]
\end{split}
\end{equation}
and
\begin{equation}\label{appendix:variance}
\begin{split}
Var[\frac{2\sigma^2}{4}T-  \frac{2\sigma^2}{4}Q] &= \frac{4\sigma^4}{16}Var[T] + \frac{4\sigma^4}{16}Var[Q]-(\frac{4\sigma^4}{16})^2 Cov(T,Q)\\[-5mm]
&= \frac{4\sigma^4}{16}(2\times(m+2(\frac{4m\mu^2}{2\sigma^2}))) + \frac{4\sigma^4}{16}(2m)\\[-5mm]
&= m\sigma^4+2\sigma^2m\mu^2\\[-5mm]
&= m\sigma^4 + 2m\sigma^2\mu^2 +m\mu^4 -m\mu^4\\[-5mm]
&=m(\sigma^2+\mu^2)^2-m\mu^4 = \bar v\\[-5mm]
\end{split}
\end{equation}

To see the Cov(T,Q)=0, we note that ({\bf x},{\bf y})$^{T}$ $\sim MVN(\mathbf{\tilde \mu}, \mathbf{\tilde \Sigma})$ where $\mathbf{\tilde \mu}=\mu \times {\bf 1}_{2m}$ for ${\bf 1}_{2m}$ a vector of one's in ${\rm I\!R}^{2m}$ and $\mathbf{\tilde \Sigma}=diag(\mathbf{\Sigma},\mathbf{\Sigma})$ is a block matrix with $\mathbf{\Sigma}=diag(\sigma^2,...,\sigma^2)\in {\rm I\!R}^{2m \times 2m}$.  Multiplying the multivariate random vector by an appropriate matrix, P, we have {\small $(\mathbf{x_1+y_1},..,\mathbf{x_m+y_m},..,\mathbf{x_1-y_1},..,\mathbf{x_m-y_m})' \sim MVN((2\mu,..,2\mu,0,..,0)',P\mathbf{\Sigma}P') $}.  By the partitioning of the full covariance matrix, we see that $(x_1+y_1,\dots,x_m+y_m)' {\perp\!\!\!\perp}(x_1-y_1,\dots,x_m-y_m)'$.  Consider $f(\boldsymbol{x})=\frac{1}{2\sigma^2}(x_1^2+\dots+x_m^2)$.  Since f(.) is a continuous function, we have $f(x_1+y_m,\dots,x_m+y_m){\perp\!\!\!\perp} f(x_1-y_m,\dots,x_m-y_m)$.  By definition of f(.), we have $T{\perp\!\!\!\perp}Q$ which implies $Cov(T,Q)=0$.

\newpage
\subsection{Supplementary Tables}
\begin{table}[ht!]
\centering
\caption{\small Within and between functional module DWE in the major depressive disorder study.  Bolded values indicate statistically significant number of DWE between the respective functional modules}
\label{table:numDWEs}
\begin{adjustbox}{width=.85\textwidth}
\begin{tabular}{llllllllllllll}
{\Large (A) Pearson, model-free} &  &  &  &  &  &  &  &  &  &  &  &  &  \\ \hline
 & SM & CIO & AUD & DMN & MEM & VIS & FPN & SAL & SUB & VAN & DAN & UNC &  \\ \hline
SM & \textbf{55} &  &  &  &  &  &  &  &  &  &  &  &  \\
CIO & 11 & 1 &  &  &  &  &  &  &  &  &  &  &  \\
AUD & 14 & 4 & 1 &  &  &  &  &  &  &  &  &  &  \\
DMN & 36 & 22 & 21 & 42 &  &  &  &  &  &  &  &  &  \\
MEM & 6 & 0 & 0 & 7 & 0 &  &  &  &  &  &  &  &  \\
VIS & 29 & 14 & \textbf{28} & 32 & 1 & 4 &  &  &  &  &  &  &  \\
FPN & 10 & 1 & 9 & 26 & 1 & 11 & 7 &  &  &  &  &  &  \\
SAL & 11 & 2 & 12 & 15 & 1 & 4 & 9 & 2 &  &  &  &  &  \\
SUB & 20 & 3 & \textbf{12} & \textbf{6} & 5 & \textbf{63} & 2 & 1 & 2 &  &  &  &  \\
VAN & \textbf{25} & 6 & 2 & 15 & 0 & 8 & 2 & 2 & 2 & 0 &  &  &  \\
DAN & \textbf{20} & 0 & 4 & 9 & 0 & 9 & 3 & 2 & 2 & 2 & 0 &  &  \\
UNC & 22 & 10 & 3 & 43 & 5 & 13 & 11 & 9 & 9 & 4 & 2 & 5 &  \\ \hline \hline
 &  &  &  &  &  &  &  &  &  &  &  &  &  \\
{\Large (B) Pearson, model-based} &  &  &  &  &  &  &  &  &  &  &  &  &  \\ \hline
 & SM & CIO & AUD & DMN & MEM & VIS & FPN & SAL & SUB & VAN & DAN & UNC &  \\ \hline
SM & \textbf{46} &  &  &  &  &  &  &  &  &  &  &  &  \\
CIO & 9 & 1 &  &  &  &  &  &  &  &  &  &  &  \\
AUD & 12 & 4 & 1 &  &  &  &  &  &  &  &  &  &  \\
DMN & \textbf{27} & 25 & 22 & 45 &  &  &  &  &  &  &  &  &  \\
MEM & 3 & 0 & 0 & 5 & 0 &  &  &  &  &  &  &  &  \\
VIS & 29 & 14 & \textbf{27} & 26 & 4 & 3 &  &  &  &  &  &  &  \\
FPN & 9 & \textbf{0} & 7 & 30 & 1 & 12 & 7 &  &  &  &  &  &  \\
SAL & 10 & 2 & 8 & 16 & 1 & 6 & 12 & 4 &  &  &  &  &  \\
SUB & 16 & 3 & 9 & 10 & \textbf{6} & \textbf{54} & 1 & 1 & 2 &  &  &  &  \\
VAN & \textbf{23} & 5 & 2 & 15 & 0 & 6 & 2 & 2 & 2 & 0 &  &  &  \\
DAN & 17 & 0 & 2 & 9 & 0 & 14 & 3 & 3 & 1 & 2 & 0 &  &  \\
UNC & 20 & 9 & 1 & 48 & 4 & 14 & 10 & 12 & 9 & 4 & 3 & 5 &  \\ \hline \hline
 &  &  &  &  &  &  &  &  &  &  &  &  &  \\
{\Large (C) Partial, model-free} &  &  &  &  &  &  &  &  &  &  &  &  &  \\ \hline
 & SM & CIO & AUD & DMN & MEM & VIS & FPN & SAL & SUB & VAN & DAN & UNC &  \\ \hline
SM & 6 &  &  &  &  &  &  &  &  &  &  &  &  \\
CIO & 2 & 0 &  &  &  &  &  &  &  &  &  &  &  \\
AUD & 2 & 3 & 3 &  &  &  &  &  &  &  &  &  &  \\
DMN & 25 & 6 & 12 & 21 &  &  &  &  &  &  &  &  &  \\
MEM & 2 & 1 & 0 & 6 & 0 &  &  &  &  &  &  &  &  \\
VIS & 12 & 5 & 7 & 16 & 1 & 7 &  &  &  &  &  &  &  \\
FPN & 13 & 4 & 5 & 19 & 1 & 7 & 3 &  &  &  &  &  &  \\
SAL & 9 & 6 & 1 & 16 & 0 & 5 & 6 & 1 &  &  &  &  &  \\
SUB & 4 & 0 & 0 & 11 & 1 & 8 & 5 & 5 & 1 &  &  &  &  \\
VAN & 2 & 3 & 0 & 11 & 0 & 4 & 0 & 3 & 1 & 0 &  &  &  \\
DAN & 6 & 3 & 1 & 9 & 0 & 4 & 1 & 5 & 3 & 1 & 1 &  &  \\
UNC & 15 & 6 & 2 & 32 & 2 & 11 & 10 & 8 & 9 & 2 & 4 & 6 &  \\ \hline \hline
 &  &  &  &  &  &  &  &  &  &  &  &  &  \\
{\Large (D) Partial, model-based} &  &  &  &  &  &  &  &  &  &  &  &  &  \\ \hline
 & SM & CIO & AUD & DMN & MEM & VIS & FPN & SAL & SUB & VAN & DAN & UNC &  \\ \hline
SM & 9 &  &  &  &  &  &  &  &  &  &  &  &  \\
CIO & 7 & 0 &  &  &  &  &  &  &  &  &  &  &  \\
AUD & 8 & 4 & 3 &  &  &  &  &  &  &  &  &  &  \\
DMN & 47 & 15 & 17 & 36 &  &  &  &  &  &  &  &  &  \\
MEM & 3 & 1 & 1 & 9 & 0 &  &  &  &  &  &  &  &  \\
VIS & 16 & 8 & 11 & 30 & 5 & 11 &  &  &  &  &  &  &  \\
FPN & 20 & 8 & 7 & 40 & 3 & 19 & 7 &  &  &  &  &  &  \\
SAL & 14 & 10 & 2 & 22 & 1 & 9 & 11 & 4 &  &  &  &  &  \\
SUB & 7 & 4 & 2 & 18 & 2 & 12 & 8 & 8 & 1 &  &  &  &  \\
VAN & 8 & 3 & 1 & 19 & 1 & 8 & 2 & 6 & 2 & 0 &  &  &  \\
DAN & 6 & 4 & 4 & 17 & 0 & 6 & 4 & 7 & 4 & 3 & 3 &  &  \\
UNC & 30 & 13 & 7 & 49 & 5 & 22 & 16 & 10 & 13 & 7 & 7 & 10 & \\ \hline \hline
\end{tabular}
\end{adjustbox}
\end{table}
\end{appendices}

\clearpage
\begin{spacing}{.05}
\bibliography{myref}
\end{spacing}

\end{document}